\documentclass[sigconf]{acmart}
\usepackage[ruled, linesnumbered, noend]{algorithm2e} 
\usepackage{setspace}

\setcopyright{rightsretained}
\copyrightyear{2025}
\acmYear{2025}
\acmDOI{10.1145/3658617.3697634}
\acmConference[ASPDAC '25]{30th Asia and South Pacific Design Automation Conference}{January 20--23, 2025}{Tokyo, Japan}
\acmBooktitle{30th Asia and South Pacific Design Automation Conference (ASPDAC '25), January 20--23, 2025, Tokyo, Japan}
\acmISBN{979-8-4007-0635-6/25/01}

\begin{document}

\title[Sequential Printed MLP Circuits for Super-TinyML Multi-Sensory Applications]{Sequential Printed Multilayer Perceptron Circuits for Super-TinyML Multi-Sensory Applications}

\author{Gurol Saglam}
\email{guerol.saglam@kit.edu}
\affiliation{%
  \institution{Karlsruhe Institute Of Technology}
  \city{Karlsruhe}
  \country{Germany}
}

\author{Florentia Afentaki}
\email{afentaki@ceid.upatras.gr}
\affiliation{%
  \institution{University of Patras}
  \city{Patras}
  \country{Greece}
}

\author{Georgios Zervakis}
\email{zervakis@ceid.upatras.gr}
\affiliation{%
  \institution{University of Patras}
  \city{Patras}
  \country{Greece}
}

\author{Mehdi B. Tahoori}
\email{mehdi.tahoori@kit.edu}
\affiliation{%
 \institution{Karlsruhe Institute of Technology}
 \city{Karlsruhe}
 \country{Germany}
 }

\renewcommand{\shortauthors}{Saglam et al.}

\begin{abstract}
Super-TinyML aims to optimize machine learning models for deployment on ultra-low-power application domains such as wearable technologies and implants.
Such domains also require conformality, flexibility, and non-toxicity which traditional silicon-based systems cannot fulfill.
Printed Electronics (PE) offers not only these characteristics, but also cost-effective and on-demand fabrication.
However, Neural Networks (NN) with hundreds of features —often necessary for target applications— have not been feasible in PE because of its restrictions such as limited device count due to its large feature sizes.
In contrast to the state of the art using fully parallel architectures and limited to smaller classifiers, in this work we implement a super-TinyML architecture for bespoke (application-specific) NNs that surpasses the previous limits of state of the art and enables NNs with large number of parameters.
With the introduction of super-TinyML into PE technology, we address the area and power limitations through resource sharing with multi-cycle operation and neuron approximation.
This enables, for the first time, the implementation of NNs with up to $35.9\times$ more features and $65.4\times$ more coefficients than the state of the art solutions.
\end{abstract}




\maketitle

\section{Introduction}\label{sec:intro}

The integration of intelligence into everyday objects is a growing trend driven by the fourth industrial revolution and the Internet-of-Things.
Diverse application domains, from smart packaging to disposable goods, fast-moving consumer goods~(FMCG), in-situ monitoring, and low-end healthcare products require low-cost, flexible, and non-toxic solutions that traditional silicon-based systems cannot fulfill~\cite{Mubarik:MICRO:2020:printedml}.
The challenge lies in achieving ultra-low cost and conformality.
While silicon VLSI is quite cheap for high volume production, the cost of circuits that integrate intelligence into already low-cost applications like smart bandages must be taken into account, and such a cost can even be less than a cent~\cite{Bleier:ISCA:2020:printedmicro}.
These ultra-low cost requirements cannot be met by silicon VLSI systems, which have inherent limitations due to their high non-recurring engineering, fabrication, and packaging costs. 
Moreover, silicon chips are too rigid to meet the flexibility and porosity needed in these application domains~\cite{Bleier:ISCA:2020:printedmicro}.

Printed electronics~(PE) have emerged as a promising solution, offering low-cost and flexible hardware fabrication. 
PE refers to fabrication techniques using printing devices, enabling the realization of large-scale and flexible hardware at a negligible cost~\cite{Mubarik:MICRO:2020:printedml}.
While PE may have limitations in precision manufacturing and feature sizes compared to silicon VLSI systems, it excels in producing cost-effective solutions for applications requiring flexibility and porosity. 
PE typically operates at frequencies ranging from a few Hz to a few KHz and has micrometer feature sizes~\cite{Henkel:ICCAD2022:expedition}.

For silicon VLSI systems, Machine Learning~(ML) classification systems are further optimized to reduce the area overhead and power consumption. TinyML is an emerging field focusing on deploying models to low-power edge devices. Super-TinyML is an extension of this field where the focus is to enable ML on even more resource-constrained ultra-low-power edge devices. This field, with its focus on ultra-low-power edge devices, is extremely enticing for ML classification systems in PE.

ML classification systems targeted for PE~\cite{Mubarik:MICRO:2020:printedml} are typically implemented with sensor data as inputs, Analog-to-Digital Converters and a classifier circuit.
However, the realization of complex ML models, like Multilayer Perceptron~(MLP), induces a large area overhead, making them very challenging to realize in PE~\cite{Mubarik:MICRO:2020:printedml}.
A promising approach to bridge this gap and enable super-TinyML for PE is to use bespoke classifier architectures in which the model's coefficients are hardwired into the hardware description and tailors the circuits to the specific dataset and model.
However existing bespoke classifier designs for PE~\cite{Mubarik:MICRO:2020:printedml, Armeniakos:TCAD2023:cross, Armeniakos:TC2023:codesign, Kokkinis:DATE2023, Armeniakos:DATE2022:axml} are still far from super-TinyML requirements.
In~\cite{Armeniakos:TCAD2023:cross,Armeniakos:TC2023:codesign, Kokkinis:DATE2023}, the authors combine the bespoke paradigm with the well-known technique of Approximate Computing~(AxC) in order to reduce the area and power overhead of combinational MLP classifiers. 
More precisely, the authors in~\cite{Armeniakos:TCAD2023:cross} approximate the multipliers by using hardware-friendly coefficients and further netlist-pruning and voltage overscaling.
The authors in~\cite{Armeniakos:TC2023:codesign}, approximate the weights by a printed-friendly retraining and truncate the accumulators.
A comprehensive analysis of neuron minimization by approximation techniques is reported in~\cite{Kokkinis:DATE2023}, using quantization, unstructured pruning, and weight clustering.

All of the previous works focus on MLPs for small sensor-based applications due to the increasing complexity of MLPs requiring many large neurons with larger multipliers and/or larger accumulators. 
More Complex tasks with many input features explode area complexity and power consumption of the current state-of-the-art fully-parallel printed MLP designs.
The authors in~\cite{Mubarik:MICRO:2020:printedml} evaluated also sequential architectures, but followed a conventional design approach and deduced eventually that fully-parallel designs are significantly more efficient. 
However, the sequential designs of~\cite{Mubarik:MICRO:2020:printedml} use a large number of registers that are very costly in PE~\cite{Bleier:ISCA:2020:printedmicro}. 

In this work, we propose a new sequential circuit for super-TinyML customized to PE which minimizes the number of employed registers by exploiting the bespoke nature of MLPs and hardwiring the coefficients using multiplexers. In addition, we approximate the neurons to further reduce the area requirements.
As a result, compared to the fully-parallel printed MLP architectures that can support limited models (up to 21 inputs and 130 coefficients), our sequential super-TinyML circuits successfully realize much larger models (up to 753 inputs and 8505 coefficients) within the constraints of PE technology. To the best of our knowledge, there is no prior work on sequential super-TinyML architectures in PE.

\noindent
\textbf{Our novel contributions within this work are as follows}:
\begin{enumerate}
\item This is the first work that proposes a sequential super-TinyML design dedicated to PE technology and its constraints.
\item We approximate sequential classifier circuits by limiting some neurons to single-cycle computations. Using both single- and multi-cycle neurons, the proposed super-TinyML design is defined as an approximate hybrid sequential architecture. An automated framework\footnote{Our framework is available at \url{https://github.com/gurolsaglam/Sequential-Printed-MLP-Circuits-for-Super-TinyML-Multi-Sensory-Applications_ASPDAC25}} is designed to extract which neurons will be multi-cycle and single-cycle after approximation.
\item This work enables the realization of printed ML classifiers with up to 753 inputs and 8505 coefficients, $35.9\times$ and $65.4\times$ more than the current state of the art, respectively.
\end{enumerate}

\section{Background on Printed Electronics}\label{sec:background}
\begin{figure}[!t]
\centering
\includegraphics[]{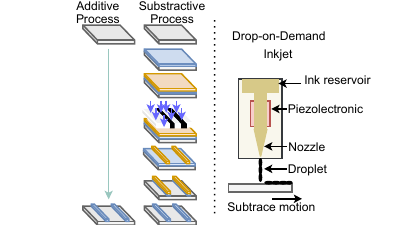}
\vspace{-3ex}
\caption{Printed Electronics fabrication techniques.}
\Description[Figure showing the different stages of printed electronics fabrication techniques, both for additive and subtractive processes.]{Figure depicts two different fabrication techniques for printed electronics. The additive process has 1 stage where the conductive material is laid on the substrate. The subtractive process has several stages; similar to the traditional lithography-based fabrication technique. The right of the figure includes a demonstration of the piezoelectric drop-on-demand inkjet printer.}
\label{fig:background} 
\vspace{-5ex}
\end{figure}

PE can be divided into two main categories of fabrications, the subtractive and the additive manufacturing processes, where the first has higher printing resolution and the second is simpler and more cost-effective.
As depicted in the left of the \figurename~\ref{fig:background}, subtractive method is similar to the traditional lithography-based fabrication technique and uses both additive and subtractive steps.
On the other hand, the additive method applies only additive steps where the functional materials are directly deposited on the substrate~\cite{Henkel:ICCAD2022:expedition}. 
The simplicity of the additive process and the fact that the technique is mask-less makes it extremely inviting for sub-cent circuits~\cite{chang2017circuits}.
However, the simplicity and the mask-less fabrication approach results in low fabrication precision.
Thus, the printed circuits have high device latency and low integration density.

In the right of the \figurename~\ref{fig:background}, the piezoelectric drop-on-demand inkjet printer is demonstrated.
The drop-on-demand mode releases ink droplets at the intended printing location. 
Piezo inkjet utilizes piezo elements to induce a shockwave through the ink, resulting in droplet ejection.
The piezoelectric type stands out due to its compatibility with various ink formulas, adjustable actuation signals, and a durable print head. 
Drop-on-demand printing is the most widely employed technique in printing functional materials, since it proves efficient in ink usage, featuring smaller ink droplet diameters ($10\text{-}50 \mu m$) and thus relatively higher printing resolution\cite{Shao:Applied_Physics2019:jet_printing}.

Furthermore, while in the past PE was mainly focused on organic transistors, nowadays transparent oxide-based inorganic printed technologies have emerged~\cite{Bleier:ISCA:2020:printedmicro}. 
Transparent oxides have one order of magnitude higher mobilities than organic semiconducting polymers.
In this work, we focus on electrolyte-gated FET~(EGFET) printed technology which, compared to the conventional inorganic channel based TFT, has smaller supply voltage ($1V$ instead of $> \! 30V$) and higher mobility values ($126 cm^2\!/\!V\!s$ instead of $\!<\!1 cm^2\!/\!V\!s$), making it suitable for battery-powered applications like wearables~\cite{Bleier:ISCA:2020:printedmicro}.

\section{Proposed Sequential Super-TinyMLs}\label{sec:framework}
In this section, we describe thoroughly the proposed architecture and the extraction of model parameters.
The sequential super-TinyML architecture is composed of controller logic, the hidden layer, the output layer, and the argmax. The hidden and output layers consist of the neurons of the MLP, where some neurons are multi-cycle and some are single-cycle.
In the first subsection, we first describe, respectively, the architecture of the multi-cycle neuron and the single-cycle neuron. 
After the description of the neuron architectures, the overall architecture and the architectural decisions to minimize the number of employed registers are described.
The second subsection consists of the high-level optimizations applied to the MLPs.
This subsection includes quantization, redundant feature pruning, and neuron approximation.

\subsection{Proposed Sequential Architecture}\label{subsec:overallarch}
\subsubsection{Multi-cycle neuron architecture}\label{subsubsec:archmulti}
\figurename~\ref{fig:neuron} depicts the state-of-the-art~\cite{Mubarik:MICRO:2020:printedml} and the proposed neuron architectures: on the left, the state-of-the-art, in the center, the multi-cycle neuron, and on the right, the single-cycle neuron. 
Typically, in a sequential neuron, registers are used to store the weights and biases. 
However, registers impose a huge area overhead in PE, and thus, their utilization should be minimized~\cite{Mubarik:MICRO:2020:printedml}. 
With the low-cost fabrication of PE technology, the weights can be hardwired in bespoke circuits, as in~\cite{Armeniakos:DATE2022:axml}. 
Using the $state$ signal provided from the controller, a multiplexer is implemented to switch, at each cycle, between the different weights.
As it will be discussed in Section~\ref{subsec:quantization}, all MLPs used in this work consider power-of-2 weights.
Thus, the weights are stored as their powers $p_i$ and their signs $s_i$ such that $w_i\!=\!s\!\times\!2^{p_i}\!$ to further reduce the area of the multiplexers.
Thus, a barrel shifter is utilized for the multiplication, with the power $p_i$ as input.
The shifted input is connected to a multiplexer with and without inverters to subtract the product from the sum when the weight is negative $s_i\!=\!1$. 
The neuron's sum is stored in a register that is reset to the bias value at the beginning of every new inference, i.e., $reset\!=\!1$.

\subsubsection{Single-cycle neuron architecture}\label{subsubsec:archsingle}
\begin{figure}[!t]
\centering
\includegraphics[width=3.35in]{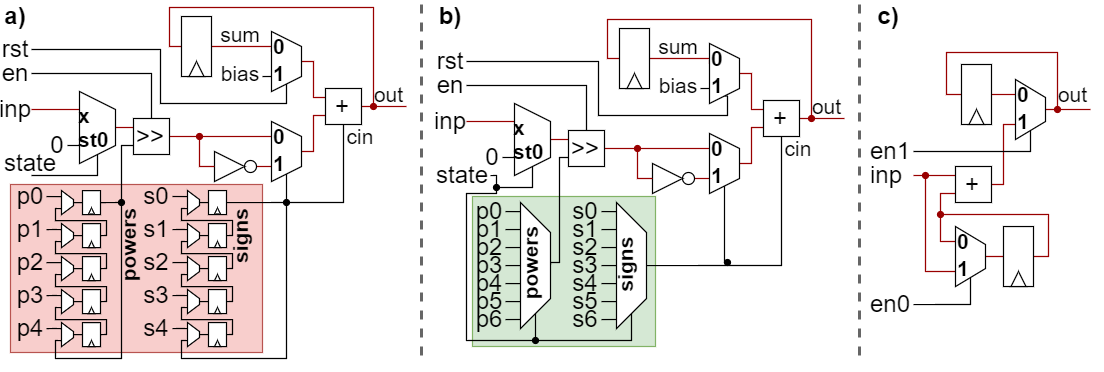} 
\vspace{-6ex}
\caption{Neuron architectures of a)~the state-of-the-art~\cite{Mubarik:MICRO:2020:printedml} and proposed: b)~multi-cycle neuron, c)~single-cycle neuron.}
\Description[Figure showing the proposed neuron architectures.]{The left of the figure shows the architecture of a single neuron. The input signal arrives at the multiplexer controlled by the state signal, then with an enable signal the input is shifted by the power of weights coming from another multiplexer controlled by the state signal. The shifted result is then added to or subtracted from the sum with an adder. The sum is kept in a register. The right of the figure has only 2 registers where 1 is used to keep the first important input until the second one arrives. Then, they are added with a 1 bit adder and the result is stored in the other register until a new inference needs to be done.}
\label{fig:neuron} 
\vspace{-4ex}
\end{figure}

The single-cycle neuron architecture is depicted on the right of \figurename~\ref{fig:neuron}.
As it is discussed in Section~\ref{subsec:neuron_approx}, based on an offline statistical analysis, we find the two most-important inputs per neuron and their corresponding expected leading-$1$ position using the training dataset and the absolute weights of each neuron.
Two enable signals, $en0$ and $en1$, are used to signify that an important input has arrived.
When $en0\!=\!1$, the neuron's first most-important input has arrived, and the expected leading-$1$ bit is stored in a 1-bit register.
When $en1\!=\!1$, the neuron's second most-important input has arrived and is directly connected to a $1$-bit adder alongside the $1$-bit that was stored previously.
The accumulation result is rewired according to the position of the leading-$1$.
This step is essential to realign the approximated results with the multi-cycle neurons of the same layer.
With this approximation, the results of the single-cycle neurons may become available earlier than the multi-cycled neurons of the same layer.

\subsubsection{Overview of the overall Super-TinyML architecture} \label{subsubsec:archoverall}
\begin{figure}[!t]
\centering
\includegraphics[scale=0.228]{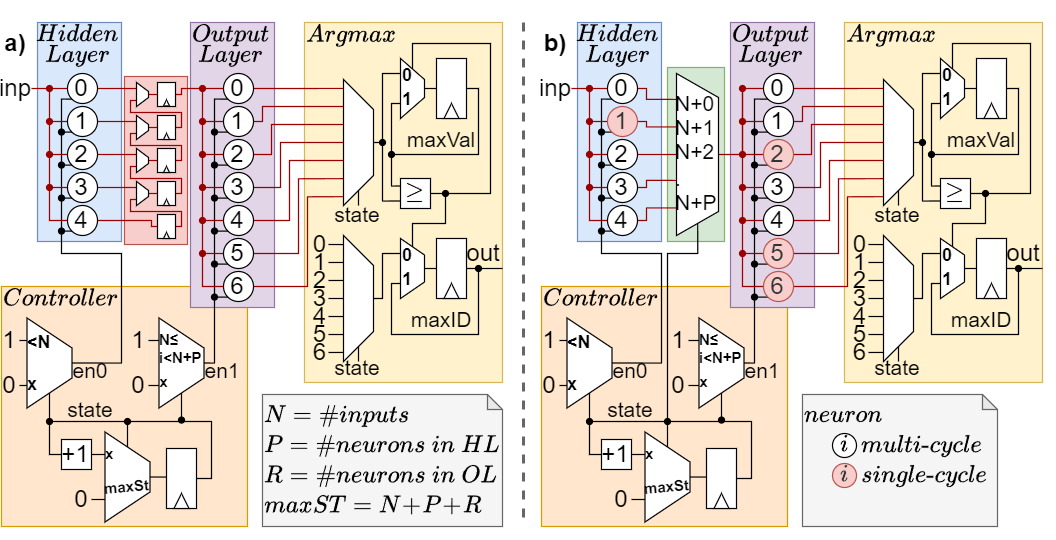} 
\vspace{-6ex}
\caption{Overview of the sequential super-TinyML architectures: a)~state-of-the-art~\cite{Mubarik:MICRO:2020:printedml}, b)~proposed.}
\Description[The overview of the sequential super-TinyML architecture is being depicted.]{The figure includes 1 Hidden Layer and 1 Output Layer, with a multiplexer between them. The multiplexer is used to switch between the different results from the Hidden Layer and to forward them to the Output Layer. The Output Layer is also connected to the Argmax where the maximum value is found and the index of the maximum value is given as the output. The figure also includes a Controller block which has a simple counter mechanism to control every block in the architecture.}
\label{fig:architecture}
\vspace{-4ex}
\end{figure}

\figurename~\ref{fig:architecture} depicts the state-of-the-art and the proposed architectures of the sequential super-TinyML circuit.
The controller logic utilizes a simple counter state machine to enable the different layers when the state value is within certain ranges ($0\!\le\!state\!\le\!N$ for hidden layer and $N\!\le\!state\!\le\!N\!+\!P$ for output layer). 
The layers also use a reset signal generated in this logic for each inference.
The state signal is also connected to the multiplexers between layers, in neurons and in argmax.
In the state-of-the-art, the layers are connected using shifting registers instead.
The proposed architecture assumes that there is only one input per cycle for the first $N$ cycles, so only one ADC is active per cycle (e.g., the rest can be power-gated).

The results of each neuron are stored in registers dedicated for accumulation.
When all the inputs are fed into the system, all of the neurons in the hidden layer will finish their calculations, and the state machine will produce an enable signal for the output layer to start the computations.
In the output layer, the same architecture is repeated and the same logic is followed.
When the results of the output layer are ready, the argmax circuit decides the winning class. 
As can be seen in \figurename~\ref{fig:architecture}, the argmax is realized with a single comparator that compares the incoming values sequentially.
There are two branches; the top branch has the values of the output layer as an input and compares them to the last largest output, while the bottom branch stores the class with the highest value.

\subsubsection{Architectural Choices}\label{subsubsec:archchoices}
\begin{figure}[!t]
\centering
\includegraphics[scale=0.8]{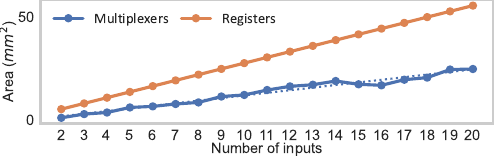}
\vspace{-2ex}
\caption{Area comparison of registers vs. multiplexers.}
\Description[Figure shows a plot of the areas of multiplexers versus registers with increasing number of inputs.]{Figure shows a plot of the areas of multiplexers versus registers with increasing number of inputs.}
\label{fig:shift_vs_mux}
\vspace{-5ex}
\end{figure}

Conventional architecture of a sequential super-TinyML includes a controller logic for synchronizing the overall execution, i.e., dataflow between the layers and the argmax.
To synchronize the data between the layers in conventional sequential MLP architectures, shifting registers are inserted between each layer, where the results of the preceding layer are stored and shifted.
Additionally, in a sequential neuron, shifting registers are used to store the weights~\cite{Mubarik:MICRO:2020:printedml}. 
However, in the proposed architecture, multiplexers are used instead of the shifting registers to achieve the same goal with much reduced area and power.
This is due to the large area overhead of the registers; the area of the shifting registers scales up linearly with the increasing number of registers required to store the output values from the preceding layer. 
Using a 2x1 multiplexer instead of 2 single-bit shifting registers already has less area ($1\!:\!4$ ratio), which increases by a smaller constant than the shifting registers. 
\figurename~\ref{fig:shift_vs_mux} presents an area comparison between the shifting registers and multiplexers w.r.t. their number of inputs.
We can observe that the shifting registers have larger area than the multiplexers for the same number of inputs.
Moreover, as the number of inputs increases, the multiplexer area scales with a smaller slope than the register, leading to larger area gains.
As a reference, for Arrhythmia dataset, which has 274 features and 1160 coefficients, replacing registers with muxes results in $4.4\times$ less area.
Additionally, to further minimize the area of multiplexers, the common denominator of weights for each neuron is calculated.
The remainder of the weights are stored in the registers or multiplexers, and the common denominator is multiplied afterwards.

\subsection{Extraction of Model Parameters}\label{subsec:optimizations}
\subsubsection{Quantization}\label{subsec:quantization}
The multiplier is the largest arithmetic unit within a neuron~\cite{Armeniakos:AxDNNsurvey}.
Thus, we minimize the multiplication area by using power-of-2~(pow2) quantization and replace the multiplier circuit by a barrel shifter.
By leveraging pow2 values, i.e., \(\pm \!w\!=\!s\!\times\!2^{p_i},p_i\!\in\!{[0,1,...,n]}\) where $n$ is the weight bitwidth, multiplication is realized by simply shifting the input by the corresponding power of the weight.
As it is described in Section~\ref{subsec:overallarch}, we realize the multiplication in the proposed sequential architecture by using only one barrel shifter per neuron.
Further, only the power~$p$ and the sign~$s$ need to be stored in the neuron, thus increasing the saving both in storing elements, i.e., muxes; and arithmetic units, i.e., multipliers.
In addition, the conventional ReLU activation function yields unbounded positive outputs, resulting in wider bit-widths at the neuron's output and subsequently larger inputs to the next layer. 
Thus, we adopt quantized ReLU~(qReLU), a quantized-friendly variant of the ReLU, which truncates certain LSBs and applies saturation of the result to a maximum value, maintaining a manageable range without the need of re-quantization steps~\cite{abram:2017:qrelu}.

\subsubsection{Redundant Feature Pruning}\label{subsec:featurepruning}
The sequential architecture flattens inputs in the time domain to reuse hardware with each input, improving hardware performance while increasing inference time. 
For multi-sensory applications, using MLPs with only a few neurons in the hidden layer, the flattened inputs dominate the time domain.
Considering the target applications have a number of inputs ranging from 44 to 753, having 3-5 hidden neurons and a few output neurons is insignificant in terms of the time domain.
Moreover, as the number of sensors increases, features become more correlated and redundant.
After the quantization of the MLP, redundant feature pruning (RFP) is applied to discard the non-important features based on their statistical relevance.
RFP is typically used in more complex ML systems such as convolutional neural networks~\cite{ayinde:2019:featurepruning}; however, to reduce the area overhead, the inference time, and consequently the energy consumption, we apply RFP in MLPs.
The reduction of weights in the MLP is translated to a reduction in area and power since smaller multiplexers and adders are required.
The inference time is shortened by a clock cycle with each pruned feature.
After this process, the MLPs retain on average $81\%$ of their original features, whereas $19\%$ are removed.

As it is described in Algorithm~\ref{alg_feature_sel}, RFP uses the weights of the hidden layer in the MLP, the training dataset, and a given threshold to calculate $N$, the minimum number of inputs required to meet the threshold accuracy.
With the weights and the training dataset, the algorithm calculates the relevance of each of the inputs.
Although the distribution of inputs in the dataset indicates which inputs have more effect on the accumulation, some inputs have larger weights.
Therefore, the products for each of the inputs are used to calculate relevance and redundancy.
Algorithm~\ref{alg_feature_sel} shows how the RFP is applied.
For each input, the expected average is calculated from the training dataset, which are multiplied with the weights of each neuron in the hidden layer.
Then, the average of the absolute products is calculated for each of the inputs, and then sorted in decreasing order.
The weights in the hidden layer neurons and the training dataset is reordered by the order of the relevance of inputs.
Afterwards, the MLP is continuously evaluated by increasing the number of features kept, until the accuracy meets the threshold.
The threshold indicates the minimum desired accuracy, which is equal to the accuracy of the quantized MLP model.
Even though the aforementioned algorithm is greedy, the runtime for the largest dataset with over 700 parameters it take less than one hour.

\subsubsection{Neuron Approximation}\label{subsec:neuron_approx}
In order to further reduce power and area, neuron approximation is applied to the MLP, in accordance with the observation that neurons can withstand different degrees of approximation.
While some neurons may need the entire input sequence to have a more accurate result, some are able to successfully estimate the result of the neuron using only a pair of bits.
This pair of bits is selected based on the average expected product of each input for each neuron.
Average expected product refers to the average of the absolute products from all the samples in the training dataset, the calculation of which is provided in Equation~\ref{eq:average_expected_product}.

\begin{equation}\label{eq:average_expected_product}
  avg\_prod_{i,n}=\frac{\sum_{j=0}^{W}(E[x_i] * |w_{n,j}|)}{W}
\end{equation}

\begin{figure}[!t]
\vspace{-3ex}
\centering
\includegraphics[scale=0.8, page=3]{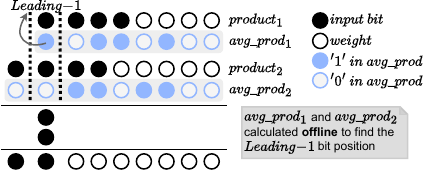}
\vspace{-3ex}
\caption{The proposed neuron approximation using the leading-1 of the average expected products.}
\Description[The figure shows an example of the proposed neuron approximation.]{The figure shows an example of how the proposed neuron approximation is done. After the two most important inputs are found for the neuron, the leading-1 index is found using the average products of each of the inputs. The leading-1 index is used for a single bit addition with a simple rewiring to correct the bit position of the approximated result.}
\label{fig:leading1} 
\vspace{-4ex}
\end{figure}

In Equation~\ref{eq:average_expected_product}, the absolute average product of each feature for each neuron is calculated, which is $avg\_prod_{i,n}$. 
$E[x_i], w_{n,j} and W$ refer to the average of feature~$x_i$, weight~$j$ of neuron~$n$, and the number of weights in neuron~$n$.
Our framework selects the two most important inputs as the inputs with the highest average expected products.
The index of the respective leading-$1$ bit is calculated from the average expected products.
Using the leading-$1$ of the two most important inputs, the result is approximated with single-bit addition.
This approximation is depicted in \figurename~\ref{fig:leading1}, where the $1$-bit addition is positioned in the Leading-$1$ column of the binary inputs.
Some neurons cause high accuracy degradation when approximated, while others have negligible impact.

\begin{algorithm}[t]
\fontsize{7.2}{8}\selectfont
\caption{Pseudocode for Redundant Feature Pruning}
\label{alg_feature_sel}
\SetKwProg{generate}{Function \emph{prune\_features}}{}{}

\generate{$MLP, x\_train, threshold$}{
$avg\_prod$=new float\text{[MLP.HL.length][x\_train.length]}\;
$avg\_x$=\text{average}$(x\_train)$\;
\For{i=0; i<MLP.HL.length; i++}{
    $avg\_prod[i]$=\text{multiply}($avg\_x, HL[i].coeffs$)\;
}
$[expected\_avg, order]$=\text{sort}(\text{average}(|$avg\_prod$|))\;
$x\_train$=$x\_train[:, order]$\;
$MLP$=\text{reorderWeights}($MLP, order$)\;

\For{i=0; i<x\_train.length; i++}{
    $acc$=$MLP$.\text{eval}($x\_train[:, 0:numofInputs]$)\;
    \If{$acc >= threshold$}{
        $N$=$numofInputs$\;
        $break$\;
    }
}
}
\end{algorithm}

We detect approximable neurons using a multi-objective genetic optimization algorithm to take advantage of its ability of effective solution space exploration.
The Non-dominated Sorting Genetic Algorithm II~(NSGA-II)~\cite{Deb:NSGA:2002} is utilized to solve this multi-objective problem.  
Each candidate solution is represented as a set of boolean values to dictate whether the neuron of the corresponding value is approximated (value 1) or not (value 0). 
Prior to the genetic optimization, an initial population biased towards solutions with mostly non-approximated neurons are generated. 
Each initial solution consists of a set of boolean values, which include only 1 approximated neuron so that the solutions chosen as the parents of following generations have higher accuracies. 
Each generation increases the number of approximated neurons until the solution with the highest number of approximable neurons and the highest accuracy is found. 
The multi-objective goal is to maximize the number of approximable neurons and the accuracy and maintain it above or equal to the desired accuracy. 
The maximum number of approximable neurons correspond to the area savings of the sequential super-TinyML circuit in an abstract manner, without the need for an extremely accurate hardware model.
The desired accuracy refers to the tolerable minimum accuracy of the MLP. 
The genetic algorithm's result is a solution, a set of boolean values, used to generate the Verilog description of the super-TinyML design.

\section{Results}\label{sec:results}
In this section, we evaluate the results of our sequential super-TinyML architecture.
Section~\ref{sec:experimental_setup} describes the experimental setup.
In Section~\ref{sec:seq_vs_sotaComb}, we compare our exact sequential design with the current state-of-the-art~\cite{Kokkinis:DATE2023}. 
In this assessment, we use the same feature selection method also on the state-of-the-art for a more fair comparison. 
In Section~\ref{sec:evaluation_approximation}, we further evaluate our sequential circuits by applying our neuron approximation. 
In Section~\ref{sec:printed_batteries_operations_and_energy}, we give a comprehensive analysis regarding energy consumption.

\subsection{Experimental Setup}\label{sec:experimental_setup}
The datasets in this work are obtained from the UCI ML repository~\cite{Dua:2019:uci}. 
In the case that non-sensor~(categorical) features exists in the datasets, we performed pre-processing to remove them. 
The datasets considered in our work are SPECTF, Arrhythmia~(Arr.), Gas Sensor~(Gas S.), Epileptic Seizure~(Epi.), Activity Recognition using Wearable Physiological Measurements~(Act.), Parkinsons~(Par.) and Human Activity Recognition~(HAR).
The MLPs are trained by following the strategy of the authors in~\cite{Mubarik:MICRO:2020:printedml, Armeniakos:TCAD2023:cross}.
The open source Qkeras~\cite{coelho:2019:Qkeras} framework is used for retraining the models for pow2 weights, while the open source PyGAD~\cite{fawzy:2023:pygad} was used for deploying the NSGA-II.
Inputs and weights are quantized to $4$-bit fixed-point and $8$-bit power-of-2 respectively, as in previous state-of-the-art~\cite{Mubarik:MICRO:2020:printedml, Armeniakos:TCAD2023:cross} except HAR dataset where the weights are quantized to $14$ bit.
The quantization resolution for HAR dataset was chosen based on the highest accuracy results after the QAT, due to the fact that the $8$-bit resolution did not reach our accuracy thresholds.
For the synthesis and simulation of the circuits, Synopsys Design Compiler T-2022.03 and VCS T-2022.06 was used respectively. 
PrimeTime U-2022.12 is used for the power analysis. 
The circuits in our work are mapped to the open-source printed EGFET library~\cite{Bleier:ISCA:2020:printedmicro}.
The combinational circuits are all synthesized at $320ms$ except SPECTF, which is synthesized at $200$ms.
The proposed sequential circuits are synthesized at different clock periods; SPECTF at $80$ms, HAR, Arr. and Gas S. at $100$ms, while the rest at $120$ms. 
The synthesis clocks are chosen based on the complexity of the MLP circuit and are in line with typical printed circuits performance~\cite{cadilha2017digital}.

\begin{figure}[t]
\centering
\includegraphics[scale=0.8]{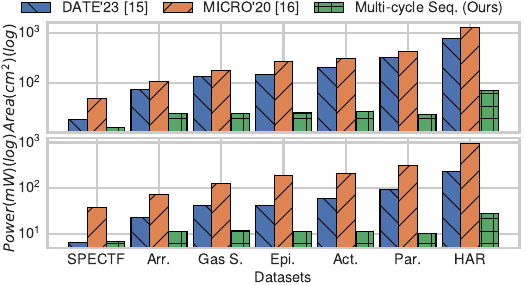}
\vspace{-3ex}
\caption{Comparison of State-of-the-Art Combinational~\cite{Kokkinis:DATE2023}, Sequential~\cite{Mubarik:MICRO:2020:printedml} and our Multi-cycle Sequential designs.}
\Description[The State-of-the-Art Combinational~\cite{Kokkinis:DATE2023}, Sequential~\cite{Mubarik:MICRO:2020:printedml} and our Multi-cycle Sequential designs are compared in this figure.]{The State-of-the-Art Combinational~\cite{Kokkinis:DATE2023}, Sequential~\cite{Mubarik:MICRO:2020:printedml} and our Multi-cycle Sequential designs are compared in this figure. The comparison is in three different parameters; area, power, and energy. It can be seen that the combinational design has less area, uses less power and energy than the Sequential, however, our Multi-cycle Sequential design has significantly less area and uses less power than the combinational, while not being able to compete with the combinational in terms of energy.}
\label{fig:comb_vs_seq} 
\vspace{-1ex}
\end{figure}
\begin{table}
\fontsize{7.2}{8}\selectfont
\setlength\tabcolsep{3pt}
  \vspace{-2ex}
  \caption{Evaluation of Accuracy, Area and Power}
  \vspace{-3ex}
  \label{tab:feataccarea}
  \renewcommand{\arraystretch}{1.0}
      \begin{tabular}{c|c|cc|cc}

    \toprule
     &  & \multicolumn{2}{c}{MICRO'20 \cite{Mubarik:MICRO:2020:printedml}} & \multicolumn{2}{|c}{Our Multi-cycle Seq.} \\
    Dataset & Accuracy & Area ($cm^2$) & Power ($mW$) & Area Gain & Power Gain\\
    \midrule
    SPECTF & 87.5 & 48.2 & 37.7 & 3.8$\times$ & 5.5$\times$ \\
    Arr. & 61.8 & 106.7 & 71.1 & 4.4$\times$ & 6.5$\times$ \\
    Gas S. & 90.7 & 182.1 & 128.9 & 7.3$\times$ & 10.9$\times$ \\
    Epi. & 93.5 & 275.8 & 187.8 & 11.0$\times$ & 16.5$\times$ \\
    Act. & 80.5 & 313.0 & 209.0 & 11.7$\times$ & 18.7$\times$ \\
    Par. & 85.5 & 437.1 & 317.4 & 18.5$\times$ & 31.1$\times$ \\
    HAR & 96.9 & 1276.2 & 969.2 & 18.1$\times$ & 34.3$\times$ \\
    \bottomrule
\end{tabular}
\vspace{-5ex}
\end{table}

\subsection{Evaluation of the Sequential Super-TinyMLs}
\subsubsection{Sequential vs State-of-the-Art}\label{sec:seq_vs_sotaComb}
\figurename~\ref{fig:comb_vs_seq} depicts the comparison between the combinational state-of-the-art in~\cite{Kokkinis:DATE2023}, the sequential state-of-the-art in~\cite{Mubarik:MICRO:2020:printedml} and our multi-cycle sequential super-TinyML design, respectively, when QAT and RFP are applied.
The results of the latter two designs are also provided with the accuracies of the models in Table~\ref{tab:feataccarea}.
The datasets in the figure are ordered by the number of coefficients in each MLP model.
In this work we focus on large multi-sensory inputs, that none of the previous works consider.
To this end, we replicate the architectures used in the previous works on our MLP models.
We compare our architecture with the combinational \cite{Kokkinis:DATE2023} that includes QAT; however, we also apply our RFP for a more fair comparison.
We also compare our multi-cycle sequential design with the conventional sequential in~\cite{Mubarik:MICRO:2020:printedml}, indicating the effect of our architectural choices.
As shown in \figurename~\ref{fig:comb_vs_seq},~\cite{Mubarik:MICRO:2020:printedml} compared to~\cite{Kokkinis:DATE2023} is larger in area and consumes more power, $1.7\times$ and $4.0\times$ on average, respectively.
Compared to~\cite{Mubarik:MICRO:2020:printedml}, our multi-cycle sequential design is $10.7\times$ smaller in area and consumes $17.6\times$ less power.
Additionally, the proposed sequential design has less area and power consumption compared to~\cite{Kokkinis:DATE2023}.
The proposed exact sequential architecture achieves on average $6.9\times$ and $4.7\times$, area and power gains over the combinational~\cite{Kokkinis:DATE2023}.
The area and power gains of our design, compared to combinational~\cite{Kokkinis:DATE2023}, range from $1.5\times$ to $13.8\times$ and from $0.9\times$ to $9.2\times$ respectively.
As depicted in \figurename~\ref{fig:comb_vs_seq}, the power of the SPECTF dataset is slightly increased.
The additional components that a sequential design requires such as registers and muxes increase the required area and power, and create an overhead for smaller datasets when compared to the combinational design.
In other words, the sequential design proves superior to its combinational counterpart, in terms of area and power, as number of inputs and subsequently the size of the MLP scale.
Since the SPECTF dataset is the smallest of the datasets, the sequential architecture's results are closer to the state-of-the-art combinational design.
Additionally, the registers consume more power in ratio to other logic gates than they occupy more area.
The area in SPECTF is $1.5\times$ smaller while the power consumption is $1.1\times$ more in the proposed design, compared to~\cite{Kokkinis:DATE2023}.

\subsubsection{Evaluation of Neuron Approximation} \label{sec:evaluation_approximation}
\begin{figure}[!t]
\centering
\includegraphics[scale=0.8]{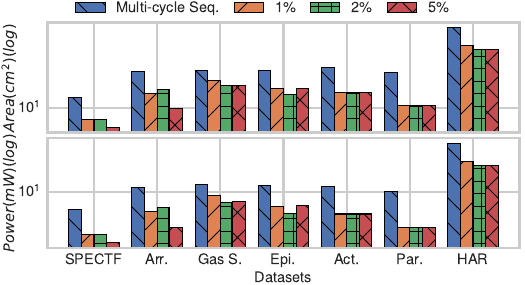}
\vspace{-3ex}
\caption{Impact of Neuron Approximation on the Hybrid design compared to the Multi-cycle Sequential design.}
\Description[The impact of Neuron Approximation on the Hybrid design compared to the Multi-cycle Sequential design are shown in this figure.]{The impact of Neuron Approximation on the Hybrid design compared to the Multi-cycle Sequential design are shown in this figure. The comparison is in three different parameters; area, power, and energy. It can be seen that with each accuracy drop threshold, the design uses less area, requires less power and energy.}
\label{fig:bar_approximation}
\vspace{-4ex}
\end{figure}

The results of the neuron approximation are depicted in \figurename~\ref{fig:bar_approximation}, highlighting the effectiveness of the proposed Neuron approximation technique using hybrid sequential architecture. 
First we apply only QAT and feature selection on the sequential architecture .
Then, on top of these optimizations we further apply neuron approximation with $1$\%, $2$\%, and $5$\% accuracy drop. 
For $1$\%, $2$\%, and $5$\% the average area gains compared to multi-cycle sequential prior to Neuron Approximation are $1.7\times$, $1.8\times$ and $1.9\times$, whereas the power gains are $1.7\times$, $1.7\times$ and $1.8\times$, respectively.
The area and power gains of all 3 accuracy constraints range from $1.3\times$ to $2.4\times$ and from $1.3\times$ to $2.3\times$.

\subsection{Energy discussion}\label{sec:printed_batteries_operations_and_energy}
\begin{figure}[!t]
\centering
\includegraphics[scale=0.8]{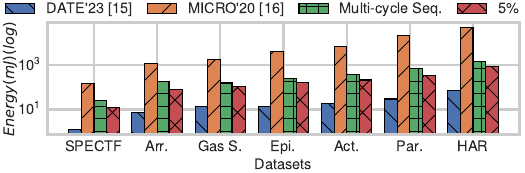}
\vspace{-3ex}
\caption{Evaluation of State-of-the-Art and our Multi-cycle Sequential designs in terms of energy.}
\Description[The energy results of the state-of-the-art and our designs are shown in this figure.]{The energy results of the state-of-the-art and our designs are shown in this figure.}
\label{fig:bar_energy}
\vspace{-4ex}
\end{figure}

The energy required for all architectures are represented in \figurename~\ref{fig:bar_energy}.
In comparison to~\cite{Kokkinis:DATE2023}, the sequential design described in~\cite{Mubarik:MICRO:2020:printedml} requires $363\times$ more energy on average, ranging from $118\times$ to $737\times$.
This steep increase is due to the large number of registers that are used.
The multi-cycle sequential architecture reduces the energy consumption to only $20\times$ compared to~\cite{Kokkinis:DATE2023}, which ranges from $12\times$ to $26\times$.
Additionally, the proposed hybrid sequential design consumes $11.5\times$ more energy compared to~\cite{Kokkinis:DATE2023}.
Our hybrid design, compared to the sequential design in~\cite{Mubarik:MICRO:2020:printedml}, demonstrates $31.6\times$ energy gains.
The sequential architecture folds the area and subsequently reduces the power in the targeted multi-sensory datasets, but unfolds in the timing domain. 
That means that each neuron in the design needs multiple cycles to calculate its result and subsequently, the sequential super-TinyML circuit makes an inference in multiple cycles compared to its combinational counterpart where only one cycle is needed.
It is important to reiterate that PE applications do not target on performance constraints.
Due to this, the energy required for the designs increases with respect to the state-of-the-art combinational architectures.
However, the primary issue in printed batteries is the the peak power they can deliver in the target applications, so even an increase in energy is tolerable~\cite{Mubarik:MICRO:2020:printedml}.

\section{Conclusion}\label{sec:conclusion}
Printed Electronics shows great potential for introducing computing and intelligence to application domains that have not yet undergone substantial amount of integration of computing, due to the unique characteristics such as conformality, flexibility, and non-toxicity.
The aforementioned characteristics are intriguing for applications like wearables, implants and in-situ monitoring. 
However, the limited device count of PE impedes the integration, while the large size of registers has deterred the realization of sequential ML architectures in the past.
Especially for machine learning classifiers, when the application uses multiple sensor inputs, the circuit gets more complex and subsequently the area and power consumption is greater.
In this work, we present a sequential super-TinyML circuit for PE that utilize both single-cycle and multi-cycle neurons and successfully.
Our evaluation shows that for MLPs with multiple input sensors, our area and power gains are on average $12.7\times$ and $8.3\times$ compared to the state-of-the-art combinational in~\cite{Kokkinis:DATE2023}.

{
\section*{Acknowledgment}
This work is partially supported by the European Research Council (ERC), and co-funded by the H.F.R.I call “Basic research Financing (Horizontal support of all Sciences)” under the National Recovery and Resilience Plan “Greece 2.0” (H.F.R.I. Project Number: 17048).
}

\begingroup
\setstretch{0.9}
\bibliographystyle{ACM-Reference-Format}
\bibliography{references}
\endgroup

\end{document}